\documentclass[12pt]{article}
\usepackage{times}
\usepackage{cite}
\usepackage{amssymb}
\usepackage{amsmath}
\usepackage{eepic}
\let\address\date 
\numberwithin{equation}{section}
\def\preprint#1#2{\vspace{-1in}\vtop{\null\hfill
\parbox[t]{1.6in}{\small\sc #1\\#2\\}}
}
\newcommand{\ads}[1]{\text{AdS}^{#1}}
\def\cf#1#2#3#4{\bibitem{#1}{#2}.~{\it #3}\,;~{#4}.}

\def\coker#1{\qopname\relax o{coKer}~{#1}}
\def\cy{Calabi--Yau~}
\def\C{{\mathbb{C}}}
\def\cp#1{{\mathbb{CP}}^{#1}}

\def\diff{\qopname\relax o{Diff}}

\def\eq#1{(\ref{#1})}
\def\eqs#1#2{(\ref{#1})--(\ref{#2})}
\def\es{{Sasaki--Einstein}~}
\def\F{{\mathcal{F}}}
\def\FI{Fayet--Iliopoulos~}
\def\G{\tilde{{\mathcal{G}}}}
\def\goth#1{{\mathfrak #1}}

\def\ie{{\em i.e.~}}
\def\ker#1{\qopname\relax o{Ker}~{#1}}
\def\map{\longmapsto}
\def\p{{\mathbb{P}}}
\def\pa{\partial}
\def\re{{\goth R\goth e}}
\def\rt{\longrightarrow}
\def\R{{\mathbb{R}}}
\def\s{{\mathbb{S}}}

\def\tilde{\widetilde}

\def\T{{\mathbb{T}}}
\def\viz{{\em viz.~}}
\def\Z{{\mathbb{Z}}}
\def\atmp#1#2#3{Adv. Theor. Math. Phys. {\bf #1}~(#2)~#3}

\def\dgga#1{dg-ga/{#1}}
\def\hepth#1{hep-th/{#1}}
\def\inv#1#2#3{Invent. Math. {\bf {#1}}~({#2})~{#3}}
\def\jdg#1#2#3{J. Diff. Geo. {\bf {#1}}~({#2})~{#3}}
\def\jhep#1#2#3{JHEP {\bf {#1}}~({#2})~{#3}}
\def\npb#1#2#3{Nucl.~Phys. {\bf B{#1}}~({#2})~{#3}}
\def\plb#1#2#3{Phys. Lett. {\bf B{#1}}~({#2})~{#3}}

\def\prl#1#2#3{Phys. Rev. Lett. {#1}~({#2})~{#3}}
\title{\preprint{ROM2F--99--14}{hep-th/~9904113} 
Candidates for anti-de~Sitter-Horizons\\\null\vfil }
\author{
Koushik Ray
\thanks{INFN--Fellow.}~\footnote{E-mail:~\tt koushik@roma2.infn.it}\\
}
\address{\begin{center} \small\sl  
Dipartimento di Fisica\\
Universit{\`a} di Roma  ``Tor Vergata'' \\
INFN  --- Sezione di Roma  ``Tor Vergata'' \\
Via della Ricerca Scientifica, 1 \\
00173  Rome  ITALY\end{center}}
\begin{document}
\maketitle\vfill
\begin{abstract}
\noindent We find, from the toric description of the moduli space
of D3--branes on non-compact six-dimensional singularities 
$\C^3/\Z_3$ and $\C^3/\Z_5$ in the blown-down
limit,  the four-dimensional 
bases on which these singular spaces 
are complex cones, and prove the existence of K\"ahler--Einstein 
metrics on these four-dimensional bases. This shows, in particular, 
that one can use 
the horizons obtained from these base spaces by a 
$U(1)$-foliation as compact parts of the target space for 
Type--IIB string theory with $\ads{5}$ in the context of the AdS-CFT 
correspondence.
\end{abstract}
\thispagestyle{empty}
\vfill\newpage
\section{Introduction}
The conjecture \cite{mal,KW} on the nexus between Type--IIB 
superstring theory on $\ads{5}\times{\cal H}$, where $\ads{5} $ is the 
five-dimensional anti-de~Sitter space and $\cal H$ is a 
five-dimensional variety, called {\em horizon} in the 
sequel, and superconformal field theories on the four-dimensional
boundary of the five-dimensional anti-de~Sitter space, thought of 
as a theory of D3--branes, has instigated a new spate of research.
The conjecture, known by now as the AdS-CFT correspondence,  
has already been tested in sundry instances. 
The tests can be broadly classified into two categories.
The first is a consideration of symmetries of the two candidates. 
Apart from a diagonal $U(1)$ subgroup of the gauge group, 
corresponding to a free-photon in the gauge theory \cite{mal,Wit-hol,MP},
the group of global isometries of the supergravity background 
in the presence of suitable four-form fluxes
corresponding to the Type--IIB superstring theory is identified with 
the R-symmetry group of the deemed superconformal theory on the 
four-dimensional boundary of $\ads{5}$ [see, for example, 
\cite{FZ}]. The other consists in an identification of operators 
in the two theories and in the comparison of their 
correlation functions [see, for example, \cite{GKP,Wit-hol,roma2}].
When the five-dimensional horizon $\cal H$ is the five-sphere, $\s^5$, 
the conformal counterpart on the D3--brane is the ${\cal N} =4$
supersymmetric gauge theory of the boundary of $\ads{5}$. 
The horizon $\cal H$ is envisaged as being 
part of the space, namely $\C^3$, transverse to the world-volume of the 
D3--brane. 

One of the interesting courses of development in the subject 
is a generalisation of the original conjecture \cite{mal}
to  models with less than 
maximal supersymmetry \cite{KW,KS,GRW,MP}.
Considering a theory of D3--branes on a Gorenstein canonical 
singularity of, for example, the type $\C^3/\Gamma$, where
$\Gamma$ is a discrete subgroup of $SU(3)$, it is possible to 
break some of the supersymmetries of the field theory of the 
brane. The corresponding candidate dual theory is then the 
Type--IIB theory compactified on $\ads{5}\times \s^5/\Gamma$
\cite{KS,KW}. For example, if $\Gamma$ is isomorphic to  a 
$\Z_2$ subgroup of an $SU(2)$ subgroup of $SU(3)$, then the 
resulting dual theory has been found to be a four-dimensional 
${\cal N} =2$ gauge theory \cite{LNV,GRW,Wit-bar}. 

Theories of a D3--brane on orbifolds of $\C^3$ of the form 
$\C^3/\Gamma$ have been considered earlier, for different groups $\Gamma$.
Examples include discrete groups $\Gamma$ isomorphic to $\Z_k$, for 
$k = 3, 5$
\cite{DGM}, 7, 9, 11\cite{muto1}, to $\Z_2\times\Z_2$ without 
discrete torsion \cite{brg,subir} and with discrete torsion 
\cite{md:dis,DF}, as well as some cases where $\Gamma$ is non-abelian
\cite{GLR,muto2}. 
In some of these  analyses \cite{DGM,muto1,brg,subir,GLR,muto2} the moduli 
space of a D3--brane at an orbifold singularity, derived as a solution
to the F- and D-flatness conditions of the corresponding 
supersymmetric gauge theory of the D3--brane, admits a toric 
geometric description. It is interesting to enquire whether one can also 
use these other cases to derive an admissible horizon for $\ads{5}$.

For an appropriate action of a $U(1)$ group 
on the horizon $\cal H$, compatible 
with the complex structure of its metric cone $C({\cal H})$, one can 
obtain the toric data of the base
${\cal H}_{B} = {\cal H}/U(1)$ from the toric data of the moduli 
space of the D3--brane on $\C^3/\Gamma$ \cite{MP}.  
It has been shown \cite{keh} that if the space ${\cal H}_B$ 
admits a K\"ahler-Einstein metric, then one can, from the knowledge 
of the $U(1)$-action, find the horizon $\cal H$ by considering 
$U(1)$-bundles of ${\cal H}_B$. 
This has been illustrated for several examples arising from partial 
resolutions of a $\C^3/(\Z_2\times\Z_2)$ singularity \cite{MP}. 

In order to obtain the horizon from the moduli space of a D3--brane, 
one starts from the field theory on the D3--brane on $\C^3/\Gamma$. This 
field theory is the blown-down limit of the theory of a D3--brane on 
the \cy singularity. In order for this theory to be dual to 
the Type--IIB
theory on $\ads{5}\times{\cal H}$, the horizon $\cal H$ is taken to 
be a space on which $\C^3/\Gamma$ is a cone. 
It has been shown that in order to retain some supersymmetry 
in the quotient theory, the horizon has to be a \es 
manifold \cite{FigF,keh}
and correspondingly, the manifold ${\cal H}_B = {\cal H}/U(1)$ must be 
K\"ahler--Einstein. If the
group-action of $U(1)$ on the horizon $\cal H$
 is \emph{regular}, meaning that the 
orbits of the $U(1)$-action are closed and have the same length, 
the base ${\cal H}_B$ of this $U(1)$-foliation 
admits a K\"ahler--Einstein metric \cite{FK,MP}. See \cite{BG} for results 
with a weaker condition of a quasi-regular $U(1)$-action.
If the discrete group $\Gamma$ is isomorphic to $\Z_3$, for example,
then the group action of $U(1)$ is regular on the horizon \cite{FK,MP}.
However, if the $U(1)$-action on the horizon is not regular,
the issue of the existence of K\"ahler--Einstein metrics 
on the corresponding bases needs be settled on a case by case basis.

In this note we consider the question of the existence of 
K\"ahler--Einstein metrics on the base ${\cal H}_B$ of the 
D3--brane moduli spaces in two examples, namely D3--branes
on  $\C^3/\Gamma$, where $\Gamma$ 
is a discrete subgroup of $SU(3)$, isomorphic to $\Z_3$ and $\Z_5$, 
that is for the orbifolds $\C^3/\Z_3$ and $\C^3/\Z_5$. 
The first corresponds to a horizon that admits a 
regular $U(1)$-action, while the other does 
not. The corresponding theories of D3-branes have been studied earlier
\cite{DGM}. In both cases we work out the toric data for the 
base, starting from the toric data of the orbifolds, obtained
earlier \cite{DGM}.
We then prove the existence of a 
K\"ahler--Einstein  metric on the respective four-dimensional bases. 
For the base of $\C^3/\Z_3$, the 
canonical toric metric is the Fubini-Study metric on $\cp{2}$, which
is {K\"ahler--Einstein}. The canonical toric metric on the base of 
$\C^3/\Z_5$ is, 
however, K\"ahler, but not Einstein.  
We find out the condition for the existence of 
a K\"ahler--Einstein metric on 
${\cal H}_B$ in terms of a deformation of the canonical K\"ahler potential,
following \cite{abru,rk}. This leads to a linear
partial differential equation whose solution purveys a K\"ahler--Einstein 
metric on the corresponding base manifold ${\cal H}_B$, thereby 
establishing 
the existence of a K\"ahler--Einstein metric on ${\cal H}_B$. 

The structure of this note is as follows. In \S\ref{cone}, we
briefly review the argument establishing the necessity of a 
K\"ahler--Einstein metric on ${\cal H}_B$, following \cite{keh}.
In \S\ref{guill}, we review the necessary information
and formulas for the calculations, following \cite{gui,abru,rk}. Finally 
in \S\ref{zed3} and \S\ref{zed5}, we present the calculation 
of K\"ahler--Einstein metrics on the bases of 
$\C^3/\Z_3$ and $\C^3/\Z_5$, before concluding in \S\ref{conclusion}.
\section{The cone, the horizon \& the base}\label{cone}
Given a space $\cal H$, the \emph{metric cone} 
$C({\cal H })$ on $\cal H$ is defined by the warped product
of $\cal H$ and the half-line, $\R^+$, as $C({\cal H}) = 
{\cal H}\times_{r^2}{\R^+}$, endowed with the metric
$\langle\cdots\rangle_{C({\cal H})} = 
dr^2 + r^2\langle\cdots\rangle_{\cal H}$.
Here and below we use subscripts to metrics descriptively whenever 
convenient. 
The conical nature of $C({\cal H})$ is discerned by noting that 
there exists a group of diffeomorphisms of $C({\cal H})$, 
isomorphic to the multiplicative group of real numbers, $\R_+$,
namely, $r\map\ t\; r$, for any positive real number $t$, which 
rescales the metric as: $\langle\cdots\rangle_{C({\cal H})} \map 
t^2\langle\cdots\rangle_{C({\cal H})}$. 
Moreover, if the space $\cal H$ admits a transitive
$U(1)$-action, then one can define another space ${\cal H}_B$ by 
quotienting $\cal H$ by the $U(1)$. Thus, the space ${\cal H}_B$ 
can be written as $C({\cal H})/(\R_+\times U(1))$, 
that, in view of the decomposition $\C^{\star} = \R_+\times U(1)$, 
can be expressed as $C({\cal H})/\C^{\star}$. 
Here $\C^{\star}$ denotes the complex-plane with the origin deleted, 
that is, $\C^{\star} = \C\setminus\{0\}$, or, in other words, the 
algebraic torus.  The metric cone on $\cal H$ can
therefore be viewed as a \emph{complex cone} over ${\cal H}_B$. In the 
sequel we shall refer to the metric cone simply as a \emph{cone}, while the 
complex cone will be explicitly mentioned. Also, we shall refer to the 
space $\cal H$ as the \emph{horizon} 
whilst ${\cal H}_B$ will be referred to as the \emph{base}
(of the complex cone as well as of the horizon). 

Let us begin by briefly recalling some features of supergravity background
solutions in the presence of D3--branes preserving some supersymmetry.
The solution to the supergravity equations of motion for 
the ten-dimensional metric retaining some supersymmetry is sought in the 
following form \cite{keh}
\begin{equation}\label{oops}
ds^2 = e^{2\Upsilon(x^m)}\eta_{\mu\nu}dx^{\mu}dx^{\nu} + 
e^{-2\Upsilon(x^m)}h_{mn}dx^mdx^n\quad m,n = 1, {\ldots}, 6;\ \ 
\mu , \nu = 7, {\ldots}, 10
\end{equation}
where the ten-dimensional space-time is taken to be split into two
parts. Four of the ten directions are parallel to the 
world-volume of the D3--brane, coordinatized by 
$x^{\mu}$, ${\mu} = 7, {\ldots}, 10$, and $\eta_{\mu\nu}$
denotes the flat-metric.
The remaining six coordinates are transverse to the D3--brane are denoted by 
$x^m$, $m=1, {\ldots}, 6$. Thus, the D3--brane is a point in the 
six-dimensional space coordinatized by $x^m, m = 1, {\ldots}, 6$.
In \eq{oops}, $\Upsilon (x^m)$ is a harmonic function of the 
coordinates of the six-dimensional space transverse to the 
world-volume of the D3--brane.
The six-dimensional metric is chosen to be of the form 
\begin{equation}\label{metricone}
ds_{C({\cal H})}^2 = h_{mn}dx^mdx^n = dr^2 + r^2g_{ij}dx^idx^j, 
\quad i,j = 1, {\ldots}, 5
\end{equation}
where the five-dimensional metric $g_{ij}$ is the metric 
on the horizon $\cal H$ that appears in the compactification 
of Type--IIB supergravity on $\ads{5}\times {\cal H}$. 
Finally, the horizon is envisaged as a  $U(1)$-bundle
over a four-dimensional space  ${\cal H}_B$, with metric of the form 
\begin{equation}
ds_{\cal H}^2= g_{ij}dx^idx^j =  ds_{{\cal H}_B}^2 + 
4(d\psi + A_adx^a)^2, \quad a,b = 1,2,3,4
\end{equation}
where the metric on the four-dimensional base ${\cal H}_B$ is 
\begin{equation}\label{gab-ans}
ds_{{\cal H}_B}^2 = g_{ab}dx^adx^b \quad a,b = 1,2,3,4.
\end{equation}
Now, if the four-dimensional space ${\cal H}_B$ is a complex K\"ahler
surface, with metric $g_{ab}$, then the 1-form $A_adx^a$ is the 
$U(1)$ connection with field strength $F$ proportional to the K\"ahler 
two-form $\omega$ of ${\cal H}_B$, \viz
\begin{equation}
F = i \omega .
\end{equation}
With the above ans\"atz, the condition for the existence of a covariant 
Killing spinor in the absence of D7--branes is found to be \cite{keh}
\begin{equation}\label{normalzn}
R_{ab} = 6g_{ab},
\end{equation}
where $R_{ab}$ denotes the Ricci tensor of the metric $g_{ab}$ in 
\eq{gab-ans}. 
Thus, the question of finding out admissible background 
solutions reduces to the question of existence of 
K\"ahler--Einstein metric on the base ${\cal H}_B$. 

One can use the known four-dimensional spaces admitting 
K\"ahler--Einstein metric as  candidates 
for ${\cal H}_B$ \cite{MP}.
Some of the choices are related to different 
partial resolutions of the moduli space of  a D3--brane on an 
orbifold $\C^3/(\Z_2\times\Z_2)$, studied earlier \cite{subir,brg}.
The connections are forged by considering the toric data
of the D-brane orbifolds of $\C^3$ in the blown-down limit and then 
realizing the horizon $\cal H$ as a line-bundle over a certain
base ${\cal H}_B$ inside the orbifold viewed as the level set 
of a combination of moment maps obtained by splitting the original
one at blown-down \cite{MP}. Another interesting case is when 
the D-brane orbifold is $\C^3/\Z_3$. The corresponding  base ${\cal H}_B$ is 
$\cp{2}$. In this case (and some others studied in \cite{MP}), the horizon
$\cal H$ has the structure of a \es manifold, with 
the aforementioned $U(1)$ providing the corresponding contact structure. 
Moreover, the action of the $U(1)$ is regular,
For such cases the existence of K\"ahler--Einstein metrics on the 
base ${\cal H}_B$
has been studied earlier \cite{FK,MP}. Let us point out that many of 
the mathematical results used in our considerations are strictly 
valid for compact spaces. However, we continue to use these results 
for the non-compact spaces at hand as the compact spaces approximate 
their non-compact counterparts in a sufficiently small neighborhood of the 
singular point.  

\section{K\"ahler--Einstein metrics on toric varieties}\label{guill}
In this section we introduce the canonical toric metric \cite{gui}
and its deformation \cite{abru,rk}. We also derive the equations 
governing the deformation by demanding the resulting metric 
to be Einstein. On the other hand, the K\"ahlerity of the deformed metric 
follows from its very construction.

Let $(X,\omega)$ be a compact, connected $2d$-dimensional manifold.
Let $\tau: \T^d \rt \diff (X,\omega)$ be an effective Hamiltonian action 
of the standard $d$-torus $\T^d$.  Let ${\mu}: X\rt \R^d$ denote the 
associated moment-map and
$\Delta$ the image of $X$ on $\R^d$ under the moment map, 
$\Delta = \mu(X)\subset\R^d$. The convex polytope $\Delta$ 
is referred to as the \emph{moment polytope}. The triple $(X,\omega ,\tau)$
is determined up to isomorphism by the moment polytope \cite{gui}.
The polytope $\Delta$ in $\R^d$ is called 
\emph{Delzant} if there are $d$ edges meeting at each vertex $p$ of 
$\Delta$ and any edge meeting at $p$ can be given the form
$p+sv_i$, for $0\leq s\leq\infty$, where
$\{v_i\}$ is a basis of $\Z^d$ \cite{gui,abru}.
Conversely, one can associate a toric variety $X_{\Delta}$ with
the above properties to a Delzant polytope $\Delta$ 
in $\R^d$, such that $\Delta$ is the moment polytope of $X_{\Delta}$. 
Let $X_{\Delta} = X$ be the toric variety associated to $\Delta$ 
that is, $X_{\Delta} = {\mu}^{\star}(\Delta )$, where $\mu^{\star}$ 
denotes the pull-back of $\mu$.

The moment 
polytope $\Delta$ can be described by a set of inequalities of the form 
$\langle y, u_i\rangle \geq \lambda_i$, $i=0,1,2,{\ldots}, d-1$. Here $u_i$ 
denotes the inward-pointing normal to the $i$-th $(d-1)$-dimensional 
face of $\Delta$
and is a primitive element of the lattice $\Z^d\subset \R^d$.
The pairing $\langle~,~\rangle$ denotes the standard scalar product in 
$\R^d$ and $y$ represents a $d$-dimensional real vector.
We can thus define a set of linear maps, 
$\ell_i : \R^n\rt \R$, 
\begin{eqnarray}\label{li}
\ell_i(y) = \langle y, u_i\rangle - \lambda_i, \quad i=0,1,{\ldots}, d-1. 
\end{eqnarray}
Denoting the interior of $\Delta$ by $\Delta^{\circ}$, 
$y\in\Delta^{\circ}$, if and only if $\ell_i(y) > 0$ for all $i$.

On the open $\T^d_{\C}$-orbit in $X_{\Delta}$, associated to a Delzant 
polytope $\Delta$, the K\"ahler form $\omega$ can be written
as \cite{gui,abru}:
\begin{eqnarray}\label{om}
\omega = {i}~\pa\bar{\pa}\mu^{\star}
\left( \sum_{i=0}^{d-1}\lambda_i\ln\ell_i
+ \ell_{\infty} \right),
\end{eqnarray}
where we have defined $\ell_{\infty}$ as the sum,
\begin{eqnarray}
\ell_{\infty} = \sum_{i=0}^{d-1}\langle y, u_i\rangle .
\end{eqnarray}
The potential in the expression \eq{om} for the K\"ahler form $\omega$
is determined by the equations for faces of the 
Delzant polytope \eq{li}. If $\omega$ is a $\T^d$-invariant K\"ahler 
form on the complex torus ${\cal M} = \C^d/2{\pi}i\Z^d$, then there exists 
a function ${\cal F}(x)$, with $x = {\re} z$, $z\in\C$, on $\cal M$, 
such that 
$\omega = 2i\pa\bar{\pa}{\cal F}$. Moreover, the moment map 
$\mu:~{\cal M}\rt\R^d$ is given by 
\begin{equation}\label{mu-leg}
\mu (z) = \frac{\pa\F}{\pa x}
\end{equation}
The K\"ahler form $\omega$ can be written in terms of this new function
(K\"ahler potential) as:
\begin{eqnarray}
\omega = \frac{i}{2} \sum_{j,k=0}^{n-1} 
\frac{\pa^2{\mathcal{F}}}{\pa x_j\pa x_k} dz_j\wedge d\bar{z}_k.
\end{eqnarray}
The moment-map $\mu$ defines a Legendre transform through \eq{mu-leg} 
and thus we can define a new set of variables $y_i$, $i=0,{\ldots},d$,
\begin{eqnarray}\label{legendre}
y_i = \frac{\pa{\mathcal{F}}}{\pa x_i}, 
\end{eqnarray}
conjugate to $x_i = \re~z_i$. 
Consequently, we can define a potential $\cal G$, conjugate to 
$\cal F$ under the Legendre transform 
\eq{legendre} on $\Delta^{\circ}$, such that 
\cite{gui,abru},  
\begin{eqnarray}\label{defG}
{\mathcal{G}} = \frac{1}{2}\sum_{k=0}^{d-1}\ell_k(y)\ln\ell_k(y).
\end{eqnarray}
The inverse of the Legendre transform \eq{legendre} can be shown to
be of the form \cite{gui}:
\begin{eqnarray}\label{invleg}
x_i = \frac{\pa{\mathcal{G}}}{\pa y_i} + r_i, \quad i=0,\cdots , d-1,
\end{eqnarray}
where $r_i$ are constants. This means that up to a linear term in 
the coordinates
$y_i$, ${\mathcal{G}}$ is the K\"ahler potential Legendre-dual to 
${\mathcal{F}}$. Moreover, the $d{\times}d$ matrix
\begin{eqnarray}
{\mathcal{G}}_{ij} = \frac{\pa^2{\mathcal{G}}}{\pa y_i\pa y_j},
\end{eqnarray}
evaluated at 
$y_i=\frac{\pa{\mathcal{F}}}{\pa x_i}$ \eq{legendre}, is the inverse of the 
matrix 
\begin{eqnarray}\label{metric-f}
{\mathcal{F}}_{ij} = \frac{\pa^2{\mathcal{F}}}{\pa x_i\pa x_j}.
\end{eqnarray}

The Ricci-tensor for the metric \eq{metric-f} takes the following form:
\begin{eqnarray}\label{Ric-1}
R_{ij} &=& -\frac{1}{2}\frac{\pa^2\ln\det~\F}{\pa x_i\pa x_j}\\
\label{Ric-tensor}
&=& -\frac{1}{2}\sum_{k,l=0}^{n-1}
{\mathcal{G}}^{lj} \frac{\pa^2 {\mathcal{G}}^{ik}}{\pa y_k\pa y_l},
\end{eqnarray}
where ${\mathcal{G}}^{ij}$ denotes the inverse
of ${\mathcal{G}}_{ij}$.
Note that, $\F$ in \eq{Ric-1} ( and in \eq{curv1} below) 
denotes  the matrix $\F_{ij}$, and not
the K\"ahler potential unlike elsewhere in this note.
The Ricci-scalar for this metric is then derived by multiplying
\eq{Ric-tensor} with ${\mathcal{G}}_{ij}$, which is the inverse of 
the metric $\F_{ij}$ in the $y$ coordinates, and is 
given by \cite{abru}
\begin{eqnarray}\label{curv1}
R &=& -\frac{1}{2}\sum_{i,j=0}^{n-1}{\mathcal{F}}^{ij} 
\frac{\pa^2 \ln\det{\mathcal{F}}}{\pa x_i\pa x_j} \\\label{curvature}
&=& -\frac{1}{2}\sum_{i,j=0}^{n-1}\frac{\pa^2{\mathcal{G}}^{ij}}
{\pa y_i\pa y_j},
\end{eqnarray}
where ${\mathcal{F}}^{ij}$ denotes the inverse 
of ${\mathcal{F}}_{ij}$.
For our purposes, it will be convenient to use the matrix 
${\mathcal{G}}_{ij}$
in the coordinates $y$. One can, in principle, rewrite all
the relevant expressions in terms of the coordinates $x_i$ 
and the matrix ${\mathcal{F}}_{ij}$. 

The expressions for the K\"ahler potential \eq{om}, 
the metric \eq{metric-f} and the curvature \eq{Ric-tensor} described above
will be referred to as the \emph{canonical} ones. 
Moreover, if $\iota:X\rt\cp{n}$, denotes a projective embedding of 
$X$ in $\cp{n}$ for some $n$, then the canonical K\"ahler form $\omega$, 
given by \eq{om}, is $\T^d$-equivariantly symplectomorphic to the pull-back 
of the K\"ahler form $\omega_{\scriptscriptstyle\text{FS}}$ corresponding 
to the Fubini-Study metric on $\cp{n}$, namely
$\iota^{\star}\omega_{\scriptscriptstyle \text{FS}}$.

However, the curvature of the canonical toric metric is 
arbitrary in general. In order to obtain a metric with
prescribed --- usually constant --- curvature, as in our cases,
one may need to deform the canonical K\"ahler form to another 
one in the same K\"ahler class. This  may be effected by 
adding a function, smooth on some open subset of $\R^n$ containing
$\Delta$, to the potential ${\mathcal{G}}$, such that the Hessian of
the new potential is positive definite on $\Delta^{\circ}$ in order 
for the new potential to be K\"ahler. 
This furnishes a deformed K\"ahler 
metric in the same K\"ahler class as the canonical one.
The variety $X_{\Delta}$ is then endowed with two different K\"ahler forms 
related by a $\T^n$-equivariant symplectomorphism since
the function $f$ is non-singular.  
One then finds out the form of this extra function by demanding 
the prescribed curvature. This method was 
used in deriving the extremal K\"ahler Metric on $\C\p^2\#\C\p^2$ 
in Calabi's form \cite{abru} as well as in finding out a Ricci-flat metric 
on the resolved D--brane orbifold $\C^3/\Z_3$ \cite{rk}. 
See \cite{cao} for another approach involving the Heat equation. 
Following the approach of \cite{abru,rk}, let us as add
a function $f$ to ${\mathcal{G}}$, and define a new potential 
$\G = {\mathcal{G}} + \frac{1}{2}f$. 
Now the matrix $\G_{ij}$ corresponding to the 
potential $\G$ assumes the form 
\begin{eqnarray}
\G_{ij} = {\mathcal{G}}_{ij} + 
\frac{1}{2}\frac{\pa^2f}{\pa y_i\pa y_j}. 
\end{eqnarray}
One can then find out the K\"ahler metric by inverting 
$\G_{ij}$ and hence the curvature for this new 
metric. This will also give rise to a new $\tilde{\F}$ 
corresponding to
${\mathcal{F}}$ and also new coordinates $\tilde{x}$. What we 
propose to do next is to write down the general form of the metric 
for a function $f$ and then determine $f$ by demanding 
that the Ricci-tensor given by the formula \eq{Ric-tensor} to be 
 proportional to the new metric $\G^{ij}$. This yields a 
differential equation for the function $f$.

For the two cases considered here, the base ${\cal H}_B$ is 
two-(complex)-dimensional. Thus, we have $i=1,2$. Specialising to 
this case, the three components of $R_{ij}$ can be written as:
\begin{eqnarray}
\label{r11}
-2R_{11} &=& 
\G^{11}\left(
\frac{\pa^2\G^{11}}{\pa y_1\pa y_1} +
\frac{\pa^2\G^{12}}{\pa y_1\pa y_2} 
\right) + 
\G^{12}\left(
\frac{\pa^2\G^{11}}{\pa y_1\pa y_2} +
\frac{\pa^2\G^{12}}{\pa y_2\pa y_2} 
\right), \\ 
\label{r12}
-2R_{12} &=& 
\G^{12}\left(
\frac{\pa^2\G^{11}}{\pa y_1\pa y_1} +
\frac{\pa^2\G^{12}}{\pa y_1\pa y_2} 
\right) + 
\G^{22}\left(
\frac{\pa^2\G^{11}}{\pa y_1\pa y_2} +
\frac{\pa^2\G^{12}}{\pa y_2\pa y_2} 
\right), \\ 
-2R_{22} &=& 
\G^{22}\left(
\frac{\pa^2\G^{22}}{\pa y_2\pa y_2} +
\frac{\pa^2\G^{12}}{\pa y_1\pa y_2} 
\right) + 
\G^{12}\left(
\frac{\pa^2\G^{22}}{\pa y_1\pa y_2} +
\frac{\pa^2\G^{12}}{\pa y_1\pa y_1} 
\right). 
\end{eqnarray}
If there exists a K\"ahler--Einstein metric on ${\cal H}_B$, then 
for some choice of the deforming function the components of the 
curvature $R_{ij}$ must be equal to the components of the metric
$\G^{ij}$. We find out a solution for $f$
by imposing this condition on $R_{ij}$.
Thus, the condition for the manifold ${\cal H}_B$ to be  
K\"ahler--Einstein, \ie  
\begin{equation}\label{reqg}
R_{ij} = \Lambda\G^{ij}, 
\end{equation}
is satisfied if  
\begin{eqnarray}
\label{reqn1}
\frac{\pa^2\G^{11}}{\pa y_1\pa y_1} &+&
\frac{\pa^2\G^{12}}{\pa y_1\pa y_2} = -2\Lambda, \\
\label{reqn2}
\frac{\pa^2\G^{22}}{\pa y_2\pa y_2}&+& 
\frac{\pa^2\G^{12}}{\pa y_1\pa y_2} = -2\Lambda, \\
\label{reqn3}
\frac{\pa^2\G^{22}}{\pa y_1\pa y_2}&+& 
\frac{\pa^2\G^{12}}{\pa y_1\pa y_1} =0,\\
\label{reqn4}
\frac{\pa^2\G^{11}}{\pa y_1\pa y_2}&+& 
\frac{\pa^2\G^{12}}{\pa y_2\pa y_2} =0.
\end{eqnarray}
Let us note that the non-covariant notation for indices in 
\eq{reqg} originates from the 
fact that, abiding by common practice, we have denoted derivatives
by subscripts and the metric under consideration has been written as 
$\G^{ij} = \G_{ij}^{-1}$ in $y$ variables.
Here $\Lambda$ is a constant parameter that determines the Ricci scalar
as $R= 2\Lambda$. However, in the following we  
keep this parameter arbitrary for book-keeping.
In \S\ref{zed3} and \S\ref{zed5}, 
we use equations  \eqs{reqn1}{reqn4} to prove the
existence of a K\"ahler--Einstein metric on ${\cal H}_B$ for the 
two cases mentioned earlier.
\section{K\"ahler--Einstein metric on the base ${\cal H}_B$ of 
$\C^3/\Z_3$}\label{zed3}
In this section we consider the blown-down limit of the moduli 
space of a D3--brane on the orbifold 
$\C^3/\Z_3$. We first find out a toric 
description of the base ${\cal H}_B$ staring from the toric data 
of $\C^3/\Z_3$. Then, following the discussion in \S\ref{guill}, 
we find out the K\"ahler--Einstein metric on this base.

\begin{figure}[h]
\begin{center}
\setlength{\unitlength}{0.00083333in}
\begingroup\makeatletter\ifx\SetFigFont\undefined%
\gdef\SetFigFont#1#2#3#4#5{%
  \reset@font\fontsize{#1}{#2pt}%
  \fontfamily{#3}\fontseries{#4}\fontshape{#5}%
  \selectfont}%
\fi\endgroup%
{
\begin{picture}(3109,3000)(0,-10)
\put(1500,2625){\circle*{100}}
\put(300,225){\circle*{100}}
\put(2700,1425){\circle*{100}}
\put(1500,1425){\circle{100}}
\path(300,225)(1500,2625)(2700,1425)(300,225)
\put(1300,2800){\makebox(0,0)[lb]{\smash{{{\SetFigFont{12}{14.4}
{\rmdefault}{\mddefault}{\updefault}$(0,1)$}}}}}
\put(-100,0){\makebox(0,0)[lb]{\smash{{{\SetFigFont{12}{14.4}
{\rmdefault}{\mddefault}{\updefault}$(-1,-1)$}}}}}
\put(2800,1375){\makebox(0,0)[lb]{\smash{{{\SetFigFont{12}{14.4}
{\rmdefault}{\mddefault}{\updefault}$(1,0)$}}}}}
\put(1300,1575){\makebox(0,0)[lb]{\smash{{{\SetFigFont{12}{14.4}
{\rmdefault}{\mddefault}{\updefault}$(0,0)$}}}}}
\end{picture}
}
\end{center}
\caption{\small\sl Plot of the columns of the toric data ${\cal T}'$ for 
 $\C^3/\Z_3$: the unfilled circle is omitted from the description in
obtaining the Delzant polytope. The filled ones correspond to ${\cal T}_B$.}
\label{fig_z3}
\end{figure}
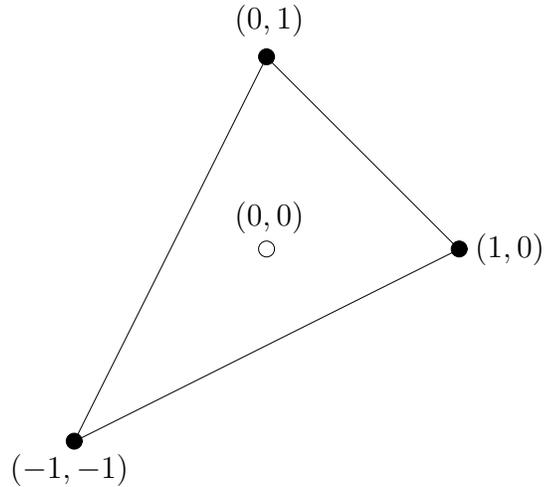

In order to obtain the base ${\cal H}_B$ for
blown-down $\C^3/\Z_3$, 
let us start from the charges of the fields.
The toric data derived from F- and D-flatness conditions of the 
gauge theory is given by \cite{DGM}
\begin{equation}\label{tordatz3}
{\cal T} = \begin{pmatrix}
-1&1&0&0\\
-1&0&1&0\\
3&0&0&1
\end{pmatrix}.
\end{equation}
The blown-down moduli space is obtained in terms of charges specified 
by the kernel of $\cal T$ with the resolution (\FI) parameter vanishing. 
The charge matrix becomes 
\begin{equation}
{\mathcal Q} = (\ker{{\cal T}})^{\text{T}} = \begin{pmatrix}
 1 & 1 & 1 & -3
\end{pmatrix},
\end{equation}
where a superscript $\scriptstyle{\text{T}}$ designates matrix transpose.
The resulting moduli space is described 
by the following moment-map equation: 
\begin{equation}\label{z3-mom}
|z_0|^2+|z_1|^2+|z_2|^2 - 3|z_3|^2 = 0,
\end{equation}
where $\{z_i|i=0,1,2,3\}$ are the homogeneous variables on the corresponding 
toric variety. The horizon $\cal H$ is obtained as a 
line-bundle from \eq{z3-mom}, following the 
construction of \cite{MP}
\begin{equation}\label{momap1}
|z_0|^2+|z_1|^2+|z_2|^2 =\zeta \qquad\text{and}\qquad  3|z_3|^2 = \zeta.
\end{equation}
The parameter $\zeta$ is left arbitrary in \eq{momap1} 
for book-keeping. It can be set to unity. 
Equations \eq{momap1} describe a space $\s^5\times\s^1$. The horizon 
$\cal H$ is obtained from \eq{momap1} after quotienting by a $U(1)$ as 
$(\s^5\times\s^1)/U(1)$, 
with the $U(1)$-action on the homogeneous variables given by 
\begin{equation}
U(1): (z_0, z_1, z_2, z_3)\map 
( e^{i\theta}z_0, e^{i\theta}z_1, e^{i\theta}z_2, e^{-3i\theta}z_3).
\end{equation}
Now, to find the toric description for the 
base ${\cal H}_B$, we collect the charges from 
\eq{momap1} into another charge-matrix 
\begin{equation}
{\mathcal Q}_B =\begin{pmatrix}
1 & 1 & 1 & 0\\
0 & 0& 0 & 3
\end{pmatrix} 
\end{equation}
and find out the toric data corresponding 
to ${\cal H}_B$ as the co-kernel of the transpose of 
${\mathcal Q}_B$. The resulting toric data is given as 
\begin{equation}\label{tordat3}
{\mathcal T}' = \coker{{\cal Q}_B^{\text{T}}} =  \begin{pmatrix}
-1&1&0&0\\
-1&0&1&0
\end{pmatrix}
\end{equation}
Leaving out the last column, this can be recognised as the toric 
data for $\cp{2}$, namely,  
\begin{equation}\label{tordatB3}
{\mathcal T}_B =  \begin{pmatrix}
-1&1&0\\
-1&0&1
\end{pmatrix}.
\end{equation}
This description corresponds to the 
``most efficient"
description of the toric variety ${\cal H}_B$, as mentioned in \cite{MP}.
The two-vectors described by the columns of ${\cal T}'$ are plotted
in Figure~\ref{fig_z3}. As in \cite{MP}, the toric data ${\cal T}_B$ in 
\eq{tordatB3} for ${\cal H}_B$ is 
obtained after omitting the point in the interior of the triangle.
Moreover, the toric data ${\cal T}'$ in
\eq{tordat3} can be identified as the first two rows in the toric data 
$\cal T$ in \eq{tordatz3} for $\C^3/\Z_3$. We shall see shortly that 
the omission of the point in the interior of the polygon 
is related to the construction of the Delzant polytope corresponding to 
 the variety.

The corresponding canonical toric metric from ${\cal T}_B$ in 
\eq{tordatB3}
is the Fubini-Study metric, as given in \cite{abru}. The Delzant 
polytope is given as 
\begin{equation}\label{poly-3}
y_1 \geq 0,\qquad y_2\geq 0 \qquad\zeta-y_1-y_2\geq 0,
\end{equation}
where we have shifted the variables $y_i$, $i=1,2$ by the respective 
values of the support functions $a_i$, $i=1,2$, at each 
of the one-dimensional cone  generators specified by the columns 
of ${\cal T}_B$, following \cite{rk}.
The K\"ahler structure
$\zeta$ is given as $\zeta=a_0+a_1+a_2$, and hence the first Chern-class
of the variety, calculated as the sum of the coefficients of $a_i$ in 
$\zeta$ \cite{gui}, is constant, $\goth{c}_1(X_{\Delta}) = 1+1+1 =3$. 
Thus, the 
variety in question is Fano. This corroborates to the fact that the 
base ${\cal H}_B$ is $\cp{2}$, as pointed out in \cite{MP}.
Let us point out that writing down the Delzant polytope provides 
a rationale for the omission of the interior points of the polygon
shown in Figure~\ref{fig_z3} (this, however, is
\emph{not} the Delzant polytope): 
they do not yield any new face for the Delzant polytope. 
The last column of \eq{tordat3} does not affect the polytope \eq{poly-3}.
The canonical potential $\mathcal G$ corresponding to the 
Delzant polytope \eq{poly-3} is, by \eq{defG},  
\begin{equation}\label{g3}
{\mathcal G} = \frac{1}{2} \biggl[ y_1\ln y_1 + y_2\ln y_2 + (\zeta -y_1-y_2)
\ln(\zeta - y_1-y_2 )\biggr].
\end{equation}
The canonical metric ensuing from the potential \eq{g3} is given by 
\begin{equation}\label{metric3}
{\mathcal G}^{ij} = \frac{2}{\zeta}\begin{pmatrix}
y_1(\zeta - y_1)&-y_1y_2
\\-y_1y_2&y_2(\zeta-y_2).
\end{pmatrix}
\end{equation}
The Ricci-tensor evaluated from \eq{metric3}, 
using \eq{Ric-tensor} can be checked to be 
proportional to the metric 
\eq{metric3}. That is, the equations  \eqs{reqn1}{reqn4} are satisfied
with $f=0$ and $\Lambda = 3/\zeta$. The Ricci-scalar for this metric is 
found from this as 
\begin{equation}\label{r-z3}
R = -\frac{1}{2}\frac{2}{\zeta}(-2-1-1-2) = \frac{6}{\zeta} = 2\Lambda, 
\end{equation}
as expected from \eq{curvature}.
Thus, we have verified  that there does exist a K\"ahler--Einstein 
metric on the base
${\cal H}_B$ of the blown-down D--brane orbifold $\C^3/\Z_3$. The metric
is diffeomorphic to the Fubini-Study metric, with a constant scalar
curvature as given in \eq{r-z3}.

A comment is in order. A Ricci-flat metric on the resolved D3--brane 
orbifold $\C^3/\Z_3$  was found out earlier using similar techniques 
in \cite{rk}. 
This needed a deformation of the canonical toric metric by a function
which was shown to be a solution of the differential equation
\footnote{The expression  \eq{fpp-sol} corrects a 
typographical error in equation (61) in \cite{rk}}
\begin{eqnarray}\label{fpp-sol}
f'' = \frac{9y_3(\xi+3y_3)^3 - (\xi+12y_3)[c + (\xi+3y_3)^3]
}{y_3(\xi+3y_3)[c + (\xi + 3y_3)^3]},
\end{eqnarray}
where $y_3$ denotes the third variable needed to form the Delzant
polytope of the three-dimensional variety, $c$ is a constant 
and $\xi$ is the resolution 
parameter (set to zero in \eq{z3-mom}).
The metric on the complex cone $C({\cal H})$
constructed from the metric \eq{metric3}  using \eq{metricone}
does not appear to be in the same form as 
the one derived in \cite{rk} for the resolved 
D--brane orbifold $\C^3/\Z_3$, with vanishing K\"ahler class.
This discrepancy, however, may be attributed to the fact that 
constancy of the scalar curvature on the base was not imposed 
on the metric in \cite{rk}, as was pointed out there.
However, since the restrictions of the metric \eq{metricone}, and the one 
derived in \cite{rk} on the base ${\cal H}_B$ are the same, with 
suitable normalisation of $y_3$, there exists some neighborhood 
of ${\cal H}_B$ in $\C^3/\Z_3$ on which 
the two metrics are diffeomorphic to one 
another, thanks to the Darboux-Weinstein theorem \cite{GS}.
\section{K\"ahler--Einstein metric on the base ${\cal H}_B$ 
of $\C^3/\Z_5$}\label{zed5}
In this section we consider the blown-down limit of the moduli space 
of a D3--brane on $\C^3/\Z_5$. First, let us find out the toric 
description of the base ${\cal H}_B$, from the toric data of the 
moduli space \cite{DGM}.

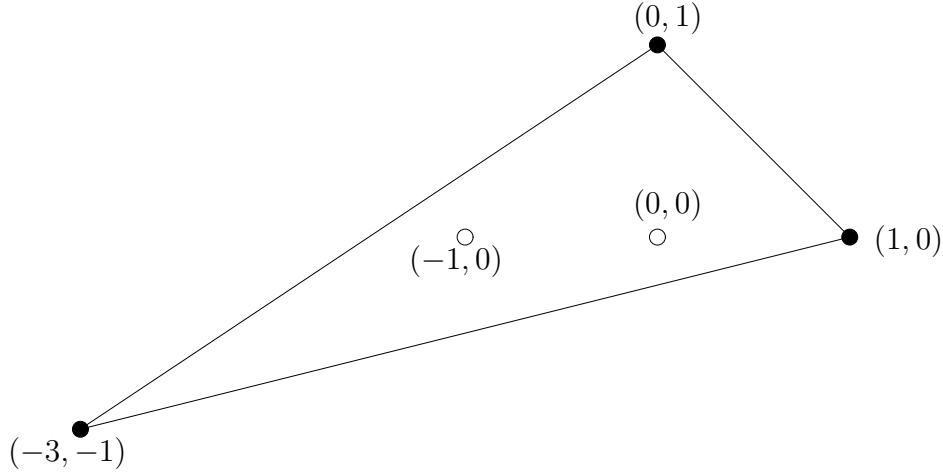
\begin{figure}[h]
\begin{center}
\setlength{\unitlength}{0.00083333in}
\begingroup\makeatletter\ifx\SetFigFont\undefined%
\gdef\SetFigFont#1#2#3#4#5{%
  \reset@font\fontsize{#1}{#2pt}%
  \fontfamily{#3}\fontseries{#4}\fontshape{#5}%
  \selectfont}%
\fi\endgroup%
{
\begin{picture}(5267,3030)(0,-10)
\put(3750,1500){\circle{100}}
\put(2550,1500){\circle{100}}
\put(3750,2700){\circle*{100}}
\put(150,300){\circle*{100}}
\put(4950,1500){\circle*{100}}
\path(150,300)(3750,2700)(4950,1500)(150,300)
\put(-300,100){\makebox(0,0)[lb]{\smash{{{\SetFigFont{12}{14.4}
{\rmdefault}{\mddefault}{\updefault}$(-3,-1)$}}}}}
\put(2200,1300){\makebox(0,0)[lb]{\smash{{{\SetFigFont{12}{14.4}
{\rmdefault}{\mddefault}{\updefault}$(-1,0)$}}}}}
\put(3600,1650){\makebox(0,0)[lb]{\smash{{{\SetFigFont{12}{14.4}
{\rmdefault}{\mddefault}{\updefault}$(0,0)$}}}}}
\put(3600,2820){\makebox(0,0)[lb]{\smash{{{\SetFigFont{12}{14.4}
{\rmdefault}{\mddefault}{\updefault}$(0,1)$}}}}}
\put(5100,1425){\makebox(0,0)[lb]{\smash{{{\SetFigFont{12}{14.4}
{\rmdefault}{\mddefault}{\updefault}$(1,0)$}}}}}
\end{picture}
}
\end{center}
\caption{\small\sl Plot of the columns of the toric data ${\cal T}'$ for 
 $\C^3/\Z_5$: the unfilled circles are omitted from the description in 
obtaining the Delzant polytope. The filled ones correspond to ${\cal T}_B$.}
\label{figz5}
\end{figure}

We shall mimic the considerations of \S\ref{zed3}. Let us start from the 
following toric data for the orbifold $\C^3/\Z_5$ \cite{DGM},
\begin{equation}
\label{t5}
{\cal T} = \begin{pmatrix}
-1 & 1 & 0 & 0& 0\\
-3 & 0 &1 &0 &-1\\
5&0&0&1&2
\end{pmatrix}.
\end{equation}
The corresponding charge matrix, evaluated as the transpose of the 
kernel of $\cal T$ takes the form
\begin{equation}\label{charge1}
\begin{pmatrix}
1&1&0&1&-3\\
0&0&1&-2&1
\end{pmatrix},
\end{equation}
that leads to two moment map equations 
in the blown-down limit with vanishing \FI parameters of the gauge theory:
\begin{eqnarray}
\label{mom5-1}
|z_0|^2 + |z_1|^2 + |z_3|^2 - 3|z_4|^2 &=& 0\\
\label{mom5-2}
|z_2|^2 - 2|z_3|^2 +|z_4|^2 &=& 0.
\end{eqnarray}
From the charge matrix \eq{charge1}, one obtains by row operations 
(changing basis) the charge matrix
\begin{eqnarray}
{\mathcal Q} = \begin{pmatrix}
1&1&3&-5&0\\
0&0&1&-2&1
\end{pmatrix}.
\end{eqnarray}
$\cal Q$ corresponds to eliminating $z_4$ from 
\eq{mom5-1} by using \eq{mom5-2} and leads to the following 
equation for the global moment map:
\begin{equation}\label{mom-5}
|z_0|^2+|z_1|^2+3|z_2|^2 -5|z_3|^2 = 0.
\end{equation}
The horizon $\cal H$ is obtained from \eq{mom-5}, or 
from the first row of $\cal Q$ constructing the line-bundle \cite{MP}: 
\begin{equation}\label{split5}
|z_0|^2+|z_1|^2+3|z_2|^2 = \zeta\qquad\text{and}\qquad 5|z_3|^2 = \zeta .
\end{equation}
Again, we have kept a free-parameter $\zeta$ for book-keeping. 
Equations \eq{split5} describe a direct product $\s^5\times\s^1$. 
The corresponding horizon is  obtained from \eq{split5} quotienting 
by a $U(1)$ as ${\mathcal H} = (\s^5\times\s^1)/{U(1)}$,
with the $U(1)$-action given by 
\begin{equation}\label{act-u1}
U(1):\, (z_0, z_1, z_2, z_3)\map 
(e^{i\theta}z_0, e^{i\theta}z_1, e^{3i\theta}z_2, e^{-5i\theta}z_3).
\end{equation}
This $U(1)$-action \eq{act-u1} is not regular. For instance,
the orbits of the points 
$(0,0,1,0)$ and $(1,0,0,0)$ are of different lengths. 
However, we proceed as before to derive the 
base ${\cal H}_B$ from this, and to this end
write the corresponding charge-matrix as
\begin{equation}
{\mathcal Q}_B = \begin{pmatrix}
1&1&3&0&0\\
0&0&0&5&0\\
0&0&1&0&1\\
0&0&0&2&0
\end{pmatrix}.
\end{equation}
This leads to the following toric data obtained as the 
co-kernel of the transpose of ${\mathcal Q}_B$, namely
\begin{equation}\label{tordat5}
{\mathcal T}' = \begin{pmatrix}
0&-3&1&0&-1\\
1&-1&0&0&0
\end{pmatrix}.
\end{equation}
From ${\mathcal T}'$, ignoring the last two columns, 
corresponding to points in the interior of the triangle, as 
shown in Figure~\ref{figz5}, we obtain the toric data for ${\cal H}_B$ as 
the following rectangular matrix:
\begin{equation}\label{tordatB5}
{\mathcal T}_B = \begin{pmatrix}
0&-3&1\\1&-1&0
\end{pmatrix}.
\end{equation}
As before, one can identify the 
matrix ${\cal T}^{\prime}$ in \eq{tordat5} inside $\cal T$ in \eq{t5}: 
${\cal T}'$ is  $\cal T$ with the last row omitted.

Having obtained the toric description for the base ${\cal H}_B$, we 
now proceed as in \S\ref{zed3}, to find out the toric metric. We 
first derive the canonical toric metric and the scalar
curvature to learn that the latter is not constant, signalling
that the canonical toric metric is not K\"ahler--Einstein. 
Following the considerations of \S\ref{guill}, we then deform the
canonical toric metric and determine the deformation 
imposing the condition that the deformed metric is {K\"ahler--Einstein},
as discussed in \S\ref{guill}.
Let us start by considering the Delzant polytope corresponding to 
the variety ${\cal H}_B$ as expressed in terms of the toric data 
${\cal T}_B$ in \eq{tordatB5}.
The Delzant polytope $\Delta$ is defined 
by the following inequalities dictated by the columns of ${\cal T}_B$:
\begin{equation}\label{delz-z5}
y_1\geq 0,\qquad y_2\geq 0, \qquad\zeta-3y_1-y_2\geq 0.
\end{equation}
Again the variables are shifted by the values of the respective support 
functions at the one-dimensional cone generators, 
 given by the columns on ${\cal T}_B$. 
This leads to  $\zeta= a_0+3a_1+a_2$, implying that the first Chern-class
$\goth{c}_1(X_{\Delta})= 1+3+1=5$, whence the variety ${\mathcal H}_B$ is 
found to be Fano. Once again the omission of the last two 
columns of ${\cal T}'$ is  justified
by their being inconsequential for constructing the Delzant polytope. 

The canonical toric metric can be obtained from the potential 
\begin{equation}\label{g5}
{\mathcal G} = \frac{1}{2} \biggl[y_1\ln y_1 + y_2\ln y_2 + 
(\zeta - 3y_1 -y_2)\ln (\zeta - 3y_1-y_2)\biggr],
\end{equation}
and the corresponding matrix ${\mathcal G}_{ij}$ is
\begin{equation}\label{non-ek}
{\mathcal G}_{ij}= \frac{1}{2A}\begin{pmatrix}
9+\frac{A}{y_1} & 3 \\
3& 1 + \frac{A}{y_2}
\end{pmatrix},
\end{equation}
where we have defined $A = \zeta -3y_1-y_2$. The metric
${\mathcal G}^{ij}$ derived by inverting \eq{non-ek}  is not Einstein. 
Indeed, the Ricci scalar for this canonical toric metric is:
\begin{equation}\label{scalar-z5}
R = \frac{26\zeta^2 + 60 \zeta y_1 + 72 y_1^2}{(\zeta  + 6y_1)^3}.
\end{equation}
The Ricci scalar $R$ is, however, positive definite, since, by 
\eq{delz-z5}, $y_1 \geq 0$ on the Delzant polytope $\Delta$.

Thus, as mentioned in \S\ref{guill}, let us now look for a deformed 
K\"ahler metric which is Einstein, by 
deforming the above potential $\mathcal G$ by a function $f$. As a 
simplifying ans\"atz, we choose $f$ to be a function of $y_1$
only, $f=f(y_1)$, since the Ricci-scalar \eq{scalar-z5} depends 
only on $y_1$. This leads to 
\begin{eqnarray}\label{g-ij}
\G_{ij} = \frac{1}{2A}\begin{pmatrix}
9 + \frac{A}{y_1} + Af'' & 3 \\
3 & 1 + \frac{A}{y_2}
\end{pmatrix},
\end{eqnarray}
where a prime signifies differentiation with respect to $y_1$.
Inverting ${\G}_{ij}$, we obtain the following expression for 
$\G^{ij}$:
\begin{eqnarray}\label{gij}
\G^{ij} = \begin{pmatrix}
\frac{2y_1}{F} (\zeta - 3y_1) & -\frac{6y_1y_2}{F} \\
-\frac{6y_1y_2}{F} & 2y_2 (1 - \frac{y_2\phi}{F})
\end{pmatrix},
\end{eqnarray}
where $F =  \zeta + 6y_1 + y_1f'' (\zeta - 3y_1)$ is a 
nowhere-vanishing function
and we have defined $\phi = (F - 9y_1)/(\zeta - 3y_1)$.
Now we solve \eqs{reqn1}{reqn4} using the expression \eq{gij}
for $\G^{ij}$. Equation \eq{reqn4} is identically satisfied 
as $f$ does not depend on $y_2$. Equations \eq{reqn2} and \eq{reqn3}
lead  to two equations respectively,
\begin{eqnarray}
\label{fineq2}
-4\left(\frac{\phi}{F}\right) -6 \left(\frac{y_1}{F}\right)' = -2\Lambda,\\
\label{fineq3}
-4\left(\frac{\phi}{F}\right)' -6 \left(\frac{y_1}{F}\right)'' = 0.
\end{eqnarray}
In deriving \eq{fineq3} from \eq{reqn3}, we have used the fact that 
$y_2\neq 0$, \ie $y\in\Delta^{\circ}$.
Equation \eq{fineq3} can be obtained by differentiating 
\eq{fineq2} with respect to 
$y_1$. Finally, from \eq{reqn1} one derives,
\begin{eqnarray}
\label{fineq1}
(\zeta - 3y_1) ( 2y_1{F'}^2 - y_1FF'') + (15y_1 - 2\zeta)FF' - 9F^2  
+\Lambda F^3 = 0,
\end{eqnarray} 
that, however, is solved identically on
using \eq{fineq2} and \eq{fineq3}. Thus, one is left with one 
single equation for $f$, namely \eq{fineq2}, that can be rewritten as
\begin{eqnarray}\label{fineq}
3y_1F' - (2\phi + 3) F + \Lambda F^2 = 0.
\end{eqnarray}
Using the expression for $\phi$, equation \eq{fineq} takes the 
following form:
\begin{equation}\label{ffin}
3 y_1 (\zeta - 3 y_1) F' - 3 (\zeta - 9 y_1) F + \biggl[ 
(\zeta - 3 y_1) \Lambda - 2 \biggr] F^2 = 0.
\end{equation}
In order to solve this, we rewrite it further in terms of a new
function $\chi = 1/F$ as:
\begin{eqnarray}\label{chieqn}
3 y_1 (\zeta - 3 y_1) {\chi}' + 3 (\zeta - 9 y_1) \chi - \biggl[ 
(\zeta - 3 y_1) \Lambda - 2 \biggr]  = 0.
\end{eqnarray}
Equation \eq{chieqn} is a linear first-order differential equation 
in $y_1$ and can be solved for $\chi$ and thence for $F$ resulting in 
\begin{eqnarray}
F = \frac{9y_1}{\Theta},
\end{eqnarray}
where $\Theta = 1 - \frac{\Lambda}{3}(\zeta - 3y_1)
+ 9c (\zeta - 3y_1)^{-2}$ and 
$c$ is a constant of integration.
One is then lead to the following equation for $f$:
\begin{equation}\label{fpp-z5}
f'' = \frac{9y_1 - (\zeta + 6y_1)\Theta}{y_1(\zeta - 3y_1)\Theta}
\end{equation}
that can be explicitly integrated to determine
the potential $\G$ in $y$-variables. 
Using \eq{fpp-z5} in the expression \eq{gij} for $\G^{ij}$ yields the 
desired K\"ahler--Einstein metric on the base ${\cal H}_B$ of the 
cone $\C^3/\Z_5$. 
\section{Conclusion}\label{conclusion}
To conclude, in this note we have demonstrated the existence 
of K\"ahler--Einstein metrics
on the bases of the complex cones obtained as moduli spaces of D3--branes
on $\C^3/\Z_3$ and $\C^3/\Z_5$ orbifolds. These cones are, in turn,
metric cones over horizons $\cal H$, and can be used as the internal 
space in  Type--IIB string theory on $\ads{5}$, in the context of 
the conjectured AdS-CFT correspondence. The analysis validates the 
candidature of the horizon obtained from the blown-down 
orbifold $\C^3/\Z_5$ along with $\C^3/\Z_3$, the latter
having already been 
pointed out in \cite{MP}. It seems possible to generalize the 
considerations in this note to establish similar results for the 
existence of K\"ahler--Einstein metrics on ${\cal H}_B$ for other 
D3--brane orbifolds
already studied in literature \cite{muto1,brg,subir,MP}. However, 
it is not clear how to extend this analysis to
cases that do not admit a toric description. 
The metrics presented here, like the one 
in \cite{rk}, are in rather special \emph{real} variables. 
It will be interesting to be able to write these metrics in terms of 
variables better-suited to studying their geometric properties.
We hope to return to this issue in future. 
\section*{\small\slshape Acknowledgements}
It is a pleasure to thank J~F Morales, S Mukhopadhyay, 
A Sagnotti and especially M Bianchi
 for illuminating discussions and useful
comments during the course 
of this work. I also thank M Abreu for a useful communication.

\end{document}